\documentclass[12pt]{article}
\usepackage{vmargin}   
\setmarginsrb{3.2cm}{2.5cm}{3.2cm}{2.5cm}{1cm}{1cm}{1cm}{1.6cm}
\newcommand{\be}{\begin{equation}}
\newcommand{\ee}{\end{equation}}
\newcommand{\bea}{\begin{eqnarray}}
\newcommand{\eea}{\end{eqnarray}}
\newcommand{\lb}{\label}
\author{David Polarski, Philippe Roche\\
Lab. de Physique Math\'ematique et Th\'eorique,\\
Universit\'e Montpellier II, UMR 5825, 34095 Montpellier Cedex 05 , France}

\begin{document}
\begin{titlepage}
\title{Can Lightcone Fluctuations be probed with   Cosmological  Backgrounds?}
\author{David Polarski\thanks{email:polarski@lpta.univ-montp2.fr},
 Philippe Roche\thanks{email:roche@lpta.univ-montp2.fr}\\
\hfill\\
LPMT,
Universit\'e Montpellier II,\\ 34095 Montpellier Cedex 05, France}
\pagestyle{plain}
\date{\today}

\maketitle

\begin{abstract}
Finding signatures of quantum gravity in cosmological observations is now actively pursued both from 
the theoretical and the experimental side. Recent work has concentrated on 
finding signatures of light-cone fluctuations in the CMB. Because in inflationary scenarios a 
Gravitational Wave Background (GWB) is always emitted much before the CMB, we can ask, in the 
hypothesis where this GWB could be observed, what is  the imprint of light cone fluctuations on 
this GWB. We show that  due to the flat nature of the GWB spectrum, the effect of lightcone 
fluctuations are negligible.
\end{abstract}

PACS Numbers: 04.60.-m, 98.80.Cq

Keywords: Quantum Gravity, Inflation, Gravitational Wave Background
\end{titlepage}

\section{Introduction}
Finding signatures of quantum gravity effects in cosmological observations is a subject which is 
currently explored from the theoretical and the experimental side \cite{Sarkar}. The existing 
approaches to quantum gravity, although still in a stage of construction (superstrings, loop 
quantum gravity), give hints on how quantum gravity effects could be detectable. 
Possible quantum gravity  effects that have been investigated, to name a few, are:

-modification of dispersion laws of propagating particles, through violation of Lorentz invariance 
(this possibility includes as well birefringence properties of the vacuum). See \cite{Jacobson} for 
a good review of experimental constraints on these effects.

-lightcone fluctuations which could have an imprint through distorsions of the spectrum of observed 
sources \cite{Ford01}.
  
We will investigate here the second possibility which has been studied in the context of the 
blackbody Cosmological Microwave Background (CMB) spectrum in \cite{Ford01}. We will review 
their work and discuss its application to the inflationary primordial Gravitational Wave 
Background (GWB).

Recent observations of the CMB anisotropy on large and small angular scales 
are in impressive agreement with the inflationary paradigm, and even with its simplest variant. 
Indeed, the latest data released by WMAP \cite{WMAP03} are in impressive agreement with a flat 
universe, a scale-invariant primordial spectrum of adiabatic fluctuations obeying gaussian 
statistics. In the inflationary paradigm, the production of fluctuations observed nowadays 
in the Universe originates from a genuine quantum process in the Early Universe. 
The simultaneous generation of a primordial GWB of quantum origin is another fundamental 
consequence of inflation. The experimental discovery of this coherent GWB spanning about 
28 orders of magnitude in frequencies would be a decisive hint at the existence of an 
inflationary stage \cite{A96}. 

The generation of quantum fluctuations could have taken place at energy scales as high as the GUT scale 
$\sim 10^{16}$ GeV but lower scales are certainly possible and perhaps even more probable. 
This is far from the Planck scale $\sim 10^{19}$ GeV, the scale at which quantum gravity effects 
are essential. Therefore if quantum gravity can have an imprint on the GWB observed today, it is not 
likely to arise through the emission spectrum of the primordial gravitons, but rather through the 
propagation of these primordial gravitons towards us. As the primordial gravitons are emitted before 
the CMB photons, actually close to the Big-bang itself, such effects due to their propagation must be 
at least as large as for CMB photons.

In our work we will discuss the effect of lightcone fluctuations through induced distortions of the 
primordial GWB. As we will show, even with conservative assumptions, requiring 
that this effect is detectable leads us to the inconsistent requirement that the 
inflation scale is of order the Planck scale. This shows that quantum gravity 
effects in the inflationary perturbations due to lightcone fluctuations are basically 
negligible, well below the threshold detection and do not affect the standard predictions. 

\section{Low Energy imprints of Quantum Gravity.}
In this section we recall two notions which have been discussed in the quantum gravity litterature: 
lightcone fluctuations and the amplification exponent.

The formulation of lightcone fluctuations is background dependent and can be expressed as follows: 
let $g^{0}_{\mu\nu}$ be a background metric and denote by $h_{\mu\nu}$ the quantum fluctuation around 
$g^{0}_{\mu\nu}.$ The full metric is $g_{\mu\nu}= g^{0}_{\mu\nu}+h_{\mu\nu},$ therefore the local 
lightcone experiences quantum fluctuations and the propagation of massive and lightlike particles 
on this space can in principle  be affected. The central question is therefore to understand  
whether these fluctuations affect the low energy physics phenomenology.

Assume that you have two spacetime events  $x,y$ and denote by $\sigma(x,y)$ one half of the squared 
geodesic distance between these points. The notion of lightcone fluctuation leads to the fact that 
$\sigma(x,y)$ is an operator whose mean value -- the classical value in the background metric --
is $\sigma_0(x,y)$ and has an uncertainty $\Delta\sigma(x,y)$. These lightcone fluctuations can in 
principle be observed: the time $t$ taken by light pulses to travel between a source and a detector 
separated by a physical distance $l$ has a quantum uncertainty $\Delta t$.

Ensuing central questions are therefore: 

1-Is the operator $\sigma(x,y)$ well defined in a theory of quantum gravity?

2-What are the situations where $\Delta  t\not=0$? How does it scale with $l$ in those cases?

3-Is this notion related to the modification of dispersion relations and does $\Delta t$ 
depend on the nature of the particles and on their energy?

The answers to these questions depend of course on the theoretical framework that one is using 
for discussing quantum gravity effects.

Question $1$ is a central question of conceptual importance but is very poorly understood in the 
quantum regime. Even in classical general relativity  $\sigma(x,y)$ is not an observable (i.e 
invariant under diffeomorphisms) and one needs to couple general relativity to matter field in 
order to build true observables \cite{Rovelli}.
 
%
%
We will not dwell further into this very interesting topic but rather state the results that have 
been obtained perturbatively in $h_{\mu\nu}$ for various backgrounds.

Writing $\sigma=\sigma_0+\sigma_1+O(h_{\mu\nu}^2)$, where $\sigma_1$ is linear in $h_{\mu\nu}$, 
it can be shown that $\Delta t=\frac{\sqrt{\langle \sigma_1^2\rangle}}{l}$ where $l$ is the 
spatial distance between the two points. 
In Minkowski space, with one spatial direction compactified on a circle of radius $R$, it was 
shown in \cite{Ford99} that 
\be
\Delta t=Ct_{P}\sqrt{l/R}~,
\ee 
where the spatial points $p,q$ are aligned in a non compactified direction, $l$ is large and $t_P$ 
denotes the Planck time. This can be seen as a refined Casimir effect which disappears in the 
decompatification limit: there are no lightcone fluctuations at first order in $h_{\mu\nu}$ 
on a Minkowski background.

Question $2$ is related to the notion of amplification exponent which is discussed in \cite{Ng}. 
Consider a rod, in a quantum theory of gravity its length is an operator with expectation value 
$l$ and a quantum uncertainty $\Delta l$. One says that  $\Delta l$ satisfies an 
amplification law with amplification exponent $\alpha$ if the following relation holds true for 
large $l$:
\begin{equation}
\Delta l = C l (l_P/l)^{\alpha}~~~~~~~~~~~~~~~~~~~~~~0<\alpha\leq 1~,\lb{ampli}
\end{equation} 
where $C$ is a constant of order unity while $l_P$ denotes the Planck length.
If the amplification law applies, then we would obtain 
\begin{equation}
\Delta t =C t_{P} (l/l_{P})^{1-\alpha}.\lb{Delal}
\end{equation} 
If it happened that $\alpha=1$, the effect would be so small that it would prevent any attempt 
to detect lightcone fluctuations.
The author of \cite{Ng} argues using two independent arguments that $\alpha=\frac{2}{3}$. 
Note that these arguments are far from being convincing and need further confirmation in a quantum 
gravity theory candidate. 

Note that this particular value of $\alpha$ lies on the edge of the range where 
fluctuations of cosmological distances are detectable. 

As we will see later, the value of $\Delta t$ is strongly constrained and for the 
propagation of photons subject to lightcone fluctuations it satisfies $\Delta t\leq 10^{-14}$. 
Taking further $l\approx H^{-1}$, we must have $\alpha > 0.5$ in order to be detectable.

As for question $3$, in the existing literature on lightcone fluctuations, $\Delta t$ is 
defined as being the quantum uncertainty of the operator $\sigma(x,y)$. 
One could measure the fluctuation of the propagation time $\Delta t$ using lightlike 
particles. However, one should bear in mind that very energetic particles could 
influence $\Delta t$ in theory of quantum gravity coupled to matter field.
In the event that these particles would experience a modification of their 
dispersion relation, $\Delta t$ could depend on the frequency $\nu$ of the 
particles used to measure their propagation time. In the present state of our 
understanding of quantum gravity coupled to matter we can only try a 
phenomenological approach and look for the possible implications of the dependance 
of $\Delta t$ on $\nu.$

\section{GWB Distortions from lightcone fluctuations}  
 
As well known, any inflationary scenario produces also a stochastic gravitational 
waves background \cite{A96}. Though its detection would represent a spectacular confirmation of 
inflation, its direct detection is unfortunately well beyond present technological 
capability. We will adopt here the point of view that the original GWB spectrum is 
distorted as a consequence of some stochastic change of frequency due 
to lightcone fluctuations, a possible quantum gravity effect. 
We want to investigate when such a distortion is significant.

We will first  assume that the time delay, and the 
corresponding change in frequency, does not depend on the frequency itself (this 
hypothesis can easily be removed) and further that the stochastic distorted frequency 
obeys a Gaussian distribution around 
the undistorted frequency. We follow here the calculations of \cite{Ford01} where it was 
assumed that the effect originates from (Gaussian) lightcone fluctuations.

We assume that as a consequence of some quantum gravity effects, a monochromatic 
gravitational wave of frequency $\nu_0$ experiences a stochastic distortion 
obeying a Gaussian distribution with dispersion 
\be
\Delta{\nu_0}= \nu_0^2 \Delta t (1- \nu_0 \Delta t)\simeq \nu_0^2 \Delta t.
\ee  
The stochasticity of $\Delta \nu_0$ comes from the gaussian random quantity 
$\Delta t$. 
Note that we first consider here a fluctuation $\Delta t$ which is {\it independent} 
of the frequency $\nu_0$. This assumption can be challenged, as we will briefly discuss 
below but models were put forward where this assumption holds, the model (\ref{Delal}) 
provides such a example. Note that we deal with models for which the inverse of the 
quantum metric fluctuations correlation time is much larger than the frequencies $\nu_0$ 
of interest to us \cite{Ford00}. 

We consider now any spectral density $F(\nu_0)$. As a result of 
the stochastic distortion of the frequencies, the new spectral density 
${\tilde F}(\nu)$ will be given by 

\bea
{\tilde F}(\nu) &=& \frac{1}{\sqrt{\pi} \Delta {\nu_0}}\int_0^{\infty} d\nu_0 ~F(\nu_0) 
                     ~e^{-\frac{(\nu_0-\nu)^2}{(\Delta {\nu_0})^2}}\\
&=& \frac{1}{\sqrt{\pi}}\int_{-\frac{\nu}{\Delta t}}^{\infty} dz
 \frac{~F(\nu+z  \Delta t)}{(\nu+z \Delta t)^2} 
                     ~e^{-\frac{z^2}{(\nu+z\Delta t)^4}}\label{expressionF}
\eea 
where we have made the change of variable $z=\frac{\nu_0-\nu}{\Delta t}$.
In order to obtain the expansion in $\Delta t$, we use the following Taylor expansion:
\begin{eqnarray}
&&F[\nu+z \Delta t]=F[\nu]+z \Delta t F'[\nu]+\frac{1}{2}z^2 \Delta t^2 F''[\nu]+o(\Delta t^2)\\
&&e^{-\frac{z^2}{(\nu+z\Delta t)^4}}=
e^{-\frac{z^2}{\nu^4}} \left( 1+\frac{4 z^3\Delta t}{\nu^5}-10 \frac{\Delta t^2 z^4}{\nu^6}+8\frac{z^6 \Delta t^2}{\nu^{10}}\right).
\end{eqnarray}
Extending the integration on $z$ in (\ref{expressionF}) up to $-\infty$ will modify it by a term 
of the type $e^{-\frac{1}{\nu 2 \Delta t^2}}$ which is negligible.
In order to compute the expansion in $\Delta t$ one is left with Gaussian integrals. 
The linear term vanishes by parity, so the calculation gives the result:
\be
\delta F \equiv {\tilde F}(\nu) - F(\nu) = F_2(\nu) \Delta t^2,\lb{tilde} 
\ee
with
\be
F_2(\nu) = 3 \nu^2 F(\nu) + 2\nu^3 F'(\nu) + \frac{\nu^4}{4} F''(\nu)~.\lb{F2}
\ee
 Therefore the fractional 
distortion of the spectral density $F$ is finally given by 
\bea
f_2(\nu) &=& \frac{{\tilde F}(\nu) - F(\nu)}{F(\nu)} \equiv \frac{F_2(\nu)}{F(\nu)} 
\Delta t^2\\
&=& (\Delta t ~\nu)^2 \Bigl[ 3 + 2\nu \frac{F'(\nu)}{F(\nu)} + 
                      \frac{\nu^2}{4} \frac{F''(\nu)}{F(\nu)} \Bigr]~.\lb{f2}
\eea 
We note that all the terms inside the brackets on the l.h.s. of (\ref{f2}) are
homogeneous. Therefore, for any spectral density which is a powerlaw in $\nu$, 
\be
F(\nu)\propto \nu^{\alpha}~~~~~~~~~~~~~~~~~~~\alpha={\rm constant}~,\lb{Falpha} 
\ee
the corresponding relative distortion $f_2(\nu)$ will be a multiple of 
$(\Delta t ~\nu)^2$, that is 
\be
f_2(\nu) = \frac{\nu^2~\Delta t^2}{4}~[\alpha^2 + 7\alpha + 12]~.\lb{f2alpha}         
\ee

This computation can be easily extended to the case where $\Delta t$ depends on the 
frequency. Indeed we can still make the change of variable 
$z=\frac{\nu_0-\nu}{\Delta t(\nu_0)}$ which leaves the equation (\ref{expressionF}) 
unchanged.
We therefore have  $z=\frac{\nu_0-\nu}{\Delta t(\nu_0)}\simeq
\frac{\nu_0-\nu}{\Delta t(\nu)}. $
As a result we still obtain the equation (\ref{f2}) with $\Delta t$ changed in 
$\Delta t(\nu).$

Let us consider the concrete case of the primordial GWB of inflationary origin. 
For example, one might be interested in the central quantity
\be
\Omega_{GW} = \frac{1}{\rho_{cr}} \frac{\partial \rho_{GW}}{\partial \ln k}~,
\ee
where the wavenumber $k$ is defined as $k\equiv \frac{2\pi}{c} \nu$. 
For a class of single-field slow-roll inflationary models, $\Omega_{GW}$ is of 
the type 
\be
\Omega_{GW}\propto k^{n_T}~,\lb{Om}
\ee
where the tensorial spectral index $n_T<0$ is constant and rather small 
(see below), hence $\Omega_{GW}(\nu)$ is a weakly decreasing function.
In that case, (\ref{f2alpha}) can be applied and one finds for the relative 
distortion of $\Omega_{GW}$ 
\be
\frac{\delta \Omega_{GW}}{\Omega_{GW}} = \frac{\nu^2~\Delta t^2}{4} 
                                    ~[n_T^2 + 7 n_T + 12]~.\lb{f2nT}
\ee
The absolute distortion $\delta \Omega_{GW}$ satisfies 
\be
\delta \Omega_{GW} \propto \nu^{2+n_T} ~\Delta t^2~.\lb{delOm}
\ee
Hence, we see that both the fractional and the absolute distortions of 
$\Omega_{GW}$ grow with $\nu$.  
Recent CMB data \cite{WMAP03} indicate that the primordial spectrum is nearly 
scale-invariant, $n_s\approx 1$. For slow-roll models with constant spectral 
indices $n_s$, $n_T=n_s-1$, the tensorial index $n_T$ is constrained by 
observations to be very small, $n_T\approx 0$, this is what is meant by a 
flat GW spectrum. For the fiducial case of a pure scale-invariant spectrum 
($n_s=1$) and $n_T=0$ we obtain 
\be
\frac{\delta \Omega_{GW}}{\Omega_{GW}} = 3 (\nu~\Delta t )^2~,\lb{f20}
\ee
while $\Omega_{GW}$ is constant. 

Note that we have for the spectral density 
$F_{GW}(\nu) \equiv \frac{d \rho_{GW}}{d \nu}\propto \nu^{n_T - 1}$ 
\bea
f_2(\nu) &=& \frac{\nu^2~\Delta t^2}{4} ~[(n_T^2 + 5n_T  + 6]\lb{f2GW}\\
\delta F_{GW} &\propto& \nu^{1+n_T} ~\Delta t^2~.\lb{delFGW}
\eea
%
%
%
As we see from (\ref{f2nT}),(\ref{f2GW}), the relative distortion for both 
quantities $\Omega_{GW},~F_{GW}$ is proportional to the quantity 
$(\nu~\Delta t)^2$.

One can also consider slow-roll models with running tensorial spectral index $n_T(k)$ 
but this will introduce no essential change, $f_2(\nu)$ will be even smaller. 
This is because $n_T$ is negative and its running is towards larger, i.e. less negative 
values. It is sufficiently illustrative for our purposes to consider a pure 
scale-invariant spectrum of density perturbations, $n_s=1$. The following result is then 
obtained instead of (\ref{f2nT}) after a straightforward calculation
\be\lb{f2k} 
\frac{\delta \Omega_{GW}}{\Omega_{GW}} = (\nu~\Delta t )^2 \Bigl[ 3 + 
      \frac{7}{4} n_{T,H} \left( 1 - n_{T,H} \ln \frac{\nu}{\nu_H} \right)^{-1}  
    + \frac{1}{2} n_{T,H}^2 \left( 1 - n_{T,H} \ln \frac{\nu}{\nu_H} \right)^{-2} \Bigr]~,
\ee
where $n_{T,H}$ is the index at, and $\nu_H$ the frequency associated to, the present 
Hubble radius. We note that (\ref{f2k}) agrees with (\ref{f2nT}) when $n_{T,H}=0$ 
because in this model there is no running when $n_T=0$. For the highest GWB 
frequencies we have $\frac{\nu}{\nu_H}\approx 10^{28}$. 

Observation of the distortion is optimal when both fractional and absolute distortions 
are maximized, which does not have to take place for the same frequencies. In addition, 
the observability is constrained by the range of frequencies where the background can be 
detected at all, whether slightly distorted or not, certainly a crucial constraint 
for the GWB.
From (\ref{f2nT},\ref{f2k}), we see immediately that the observability of the GWB distortion 
requires that $(\nu \Delta t)$ is of order one, say 10\%. As $\Delta t$ is 
exceedingly small, only very high frequencies $\nu\sim \Delta t^{-1}$ might display a 
potentially observable distortion. Unfortunately, at very high frequencies a direct 
detection is problematic at the present time.

It is unlikely that the inflationary GWB could be detected in the near future. Therefore 
it is impossible at the present time to put experimental constraints on the effect of 
lightcone fluctuations on the propagation of primordial gravitons. 
Fortunately we have one cosmological background, namely the Cosmic Microwave Background (CMB),
for which we have stringent constraints on possible distortions. Therefore, if we assume the 
existence of distortions generated by lightcone fluctuations as modeled above, we can set 
an upper bound on $\Delta t$ using (\ref{tilde},\ref{F2},\ref{f2}) as explained in \cite{Ford01}.  
The accurately measured CMB blackbody spectrum constrains the relative distortion 
to be at most of the order allowed by the FIRAS (COBE) data. 
Taking for the undistorted distribution $F_{CMB}(\nu)$ 
\be
F_{CMB}(\nu) \equiv \frac{d \rho_{CMB}}{d \nu} = 
                    \frac{8\pi h}{c^3} ~\frac{\nu^3}{e^{\frac{h\nu}{k_B T}}-1}~,\lb{FCMB}
\ee
(we have put back in (\ref{FCMB}) the velocity of light $c$) a blackbody radiation 
distribution with temperature $T\simeq 2.725{\rm K}$, one can compute its distortion 
due to lightcone fluctuations using the expressions (\ref{tilde},\ref{F2},\ref{f2}). 
As could be expected, the maximal distortion will be for frequencies close to the peak 
$\nu_{max}$ of the CMB blackbody spectrum, i.e. for 
\be
\nu\approx \nu_{max} = 1.60 \times 10^{11}~{\rm Hz}~,  
\ee
or equivalently $x\equiv \frac{h \nu}{k_B T} \approx x_{max} = 2.82$, the variable 
introduced in \cite{Ford01}. 
Note that for the CMB, the fractional distortion $f_{2,CMB}$ satisfies 
\be
f_{2,CMB} \propto \Delta t^2 ~\nu^4~~~~~~~~~~~~~~{\rm for}~\nu\to \infty~,
\ee 
and grows for large frequencies. However, due to the exponential decrease of 
the blackbody spectrum, the {\it absolute} distortion is maximal near $\nu_{max}$ 
and tends to zero for large frequencies. This is in contrast with the properties 
of the quantity $F_{GW}(\nu)$ for which both absolute and relative distortions grow 
for large frequencies.

The observed blackbody spectrum of the CMB allow for a relative distortion of order 
$\sim 10^{-5}$ near the peak of the spectrum. Plugging in the numbers, the quantity 
$\Delta t$ {\it for photons} emitted by the CMB is then found to satisfy the bound 
\be
\Delta t\leq 10^{-14} {\rm s}~. 
\ee
Though the primordial inflationary gravitational wave background is probably 
undetectable in the foreseeable future (for technological reasons), still 
something interesting can be learnt from (\ref{f2GW}).
With the optimistic assumption $\Delta t = 10^{-14}$ for primordial gravitons, 
one would need frequencies as high as $\nu\sim 10^{13}$Hz in order for 
(\ref{f2nT},\ref{f2k}) to be non negligible. This requires in turn an 
inflationary scale that is of order the Planck mass $M_p$!

Indeed, the highest frequencies of the stochastic gravitational wave background are 
obtained for those gravitational waves (tensor perturbations) leaving the Hubble 
radius (``horizon'') at the end of inflation.
Gravitational waves of the size of the present Hubble radius left the Hubble radius 
(during inflation) $N$ e-folds before the end of the inflationary stage. 
Let us consider two inflationary stages with two different energy scales 
$\Lambda_1$ and $\Lambda_2$ ($H^2\propto \Lambda^4$). Then it is not hard to deduce 
the following relation
\be
e^{N_2} \simeq \frac{\Lambda_2}{\Lambda_1} e^{N_1}~.
\ee
Even for an inflationary scale $\Lambda_1$ of order the GUT scale, $\Lambda_1\sim 10^{16}$Gev, 
$N_1\sim 65$ and the highest frequencies are $\nu_{max}\sim 10^{10}$Hz. 
For inflation at lower scales, the highest frequencies will be lower.
In order to get frequencies as high as $10^{13}$Hz one needs a $\Lambda$ of order the 
Planck scale! 
It is interesting that the upper bound $\Delta t\leq 10^{-14}{\rm s}$ coming from CMB data, 
totally independently of any model put forward for the generation of the primordial 
perturbations, implies that the inflationary scale should be close to the Planck scale in 
order for (\ref{f2nT},\ref{f2k}) to be non negligible. 

In the inflationary framework, the spectrum near the highest frequencies requires a  
careful treatment \cite{P01}. 
Indeed, these modes were barely outside the Hubble radius (``horizon''). This is 
in contrast the modes on cosmological scales which were outside the Hubble radius during 
most of the universe expansion. It is this all important property that leads to the 
effective quantum to classical transition of all primordial perturbations, including 
primordial gravitons, on cosmological scales \cite{PS96}. This will not apply 
to the high frequency end of the GWB spectrum.

Leaving aside these problems, at the Planck scale (or higher) the dominant quantum gravity 
effects are anyway expected to be seen on the emission spectrum itself rather than 
on the propagation time of the gravitons. We conjecture that  
a phase transition taking place around the end of the inflationary stage and associated with 
some feature, like a jump in the first or higher order derivatives in the inflaton potential 
\cite{P99} could alleviate the problem. However observing these potential effects would still 
remain very unlikely. 
A more encouraging possibility would be that $\Delta t$ is a growing 
function of $\nu$: from a fundamental perspective much work needs to be done on the computation 
of light cone fluctuations from a theory of quantum gravity.
A different effect has been studied in the formalism of Loop Quantum Gravity where  it has been 
stated that modification of dispersion relation of the  propagating matter (depending on the 
frequency $\nu$) could arise when evaluated in a spin network state of Loop Quantum Gravity. 
However because these states are not physical in the sense that they do not satisfy the 
hamiltonian constraint, conclusions drawn from these computations, if not wrong, have to be 
taken with the greatest care.
A non perturbative understanding of lightcone fluctuations needs to be adressed but as far as 
their observability in a cosmological background is concerned the conclusion of our work is negative.

\end{document}